\documentclass[12pt,leqno]{article} 
\usepackage[dvips]{graphics}
\usepackage  {epsfig}      
\newfont{\rams}{msbm10 scaled\magstep1}
\newfont{\iams}{msbm8}
\newfont{\gotic}{eufm10 scaled\magstep1}
\newfont{\bellap}{eusm10 scaled\magstep1}
\newfont{\bellaps}{eusm7} 
\newcommand{\rea}{\mbox{\rams \symbol{'122}}}

\newcommand{\rind}{\mbox{\iams \symbol{'122}}} 
\newcommand{\ecar}{\mbox{\gotic \symbol{'145}}}
\newcommand{\Pb}{\mbox{\bellap P}}
\newcommand{\Kb}{\hspace{1pt} \mbox{\bellap K}}
\newcommand{\np}{\mbox{\bellap N}}
\newcommand{\nps}{\mbox{\bellaps N}}
\newcommand{\nq}{{\sf n}_{q}}
\newcommand{\hw}{\hbar \omega}
\newcommand{\kB}{k_{B}}
\newcommand{\TL}{T_{L}}
\newcommand{\kT}{\kB \TL}
\newcommand{\mass}{m^{*}}
\newcommand{\Nse}{\bar{N}}
\newcommand{\aN}{\alpha \Nse}
\newcommand{\Dw}{\Delta w}
\newcommand{\dm}{\displaystyle}
\newcommand{\p}{\hspace{1pt} .}
\newcommand{\sv}{\hspace{1pt} ,}
\newcommand{\ud}{:=}
\newcommand{\freccia}{\rightarrow}
\newcommand{\perogni}{\quad \mbox{for every }}

\newcommand{\bE}{{\bf E}}
\newcommand{\alk}{\alpha_{k}}
\newcommand{\alt}{\tilde{\alpha}}
\newcommand{\bk}{{\bf k}}
\newcommand{\bka}{\tilde{{\bf k}}}
\newcommand{\bx}{{\bf x}}
\newcommand{\tx}{(t, \bx)}
\newcommand{\txk}{(t, \bx, \bk)}
\newcommand{\vk}{{\bf v}(\bk)}
\newcommand{\txu}{(t, \bx, u)}

\newcommand{\uph}{u + \hw}
\newcommand{\umh}{u - \hw}
\newcommand{\wpa}{w + \alpha}
\newcommand{\wma}{w - \alpha}
\newcommand{\wmap}{(w - \alpha)_{+}}
\newcommand{\Kkak}{K(\bka,\bk)}
\newcommand{\Kkka}{K(\bk,\bka)}
\newcommand{\Kzer}{K_{0}(\bk, \bka)}
\newcommand{\Kzerin}{K_{0}(\bka, \bk)}
\newcommand{\ftxk}{f \txk}
\newcommand{\Ntxu}{N(t, \bx, u)}
\newcommand{\bV}{{\bf V}}
\newcommand{\Vtxu}{\bV (t, \bx, u)}
\newcommand{\su}[1]{\sigma(#1)}
\newcommand{\Kup}{\Kb (u, \uph)}
\newcommand{\Kum}{\Kb (u, \umh)}
\newcommand{\en}{\varepsilon}
\newcommand{\enk}{\en(\bk)}
\newcommand{\enka}{\en(\bka)}
\newcommand{\Pk}{\Pb(\bk)}
\newcommand{\Rk}{ \rea^{3}}

\newcommand{\devp}[2]{ \frac{\partial #1}{\partial #2}}
\newcommand{\devo}[2]{ \frac{d \hspace{1pt} #1}{d \hspace{1pt} #2}}
\newcommand{\Itre}{\int_{\scriptstyle \rind^{3}}}
\newcommand{\Ik}[1]{\Itre #1 \hspace{2pt} d \bk}
\newcommand{\Ika}[1]{\Itre #1 \hspace{2pt} d \bka}
\newcommand{\Iu}[1]{\int_{u_{0}}^{+ \infty} #1 \hspace{2pt} du}
\newcommand{\Iw}[1]{\int_{0}^{+ \infty} 
                    \hspace{-10pt} #1 \hspace{1pt} dw}
\newcommand{\ddm}{\hspace{1pt} \delta (\enka - \enk - \hw) }
\newcommand{\ddp}{\hspace{1pt} \delta (\enka - \enk + \hw) }
\newcommand{\dee}{\hspace{1pt} \delta (\enka - \enk) } 
\newcommand{\dpm}{\hspace{1pt} \delta (\enka - \enk \pm \hw) }
\newcommand{\deu}{\hspace{1pt} \delta (\enk - u) }

\newcommand{\eq}[1]{\mbox{{\rm(\ref{#1})}}}
\newcounter{nsez}
\newcommand{\sez}[1]{\\[10pt] {\large \bf \addtocounter{nsez}{1}
             \noindent \thensez. #1} \\}
\newcounter{nsbs}

\newcommand{\cita}[1]{\cite{#1}} 
\newcommand{\auto}[2]{{\sc #2} {\sc #1} }
\newcommand{\arti}[5]{\hspace{-6pt}, {\it #1}, {#2}, {#3} (#5), pp. #4.}
\newcommand{\libr}[4]{\hspace{-6pt}, {\it #1}, #3, #2, #4. }
\begin{document}
\baselineskip=24pt
\jot=8pt
\arraycolsep=1pt
\begin{center}
{\large \bf A Novel Approach to Spherical Harmonics Expansion for 
Electron Transport in Semiconductors}
\\[3mm]
{\sc S.~F.~Liotta and A.~Majorana} 
\\
{Dipartimento di Matematica - University of Catania - } \\
{ Viale A.~Doria 6, 95125 Italy}
\end{center}
{\bf Abstract.}
A set of equations is derived from the Boltzmann kinetic equation 
describing charge transport in semiconductors. The unknowns of these 
equations depend on the space-time coordinates and the electron 
energy. 
The non-parabolic and parabolic band approximation are treated in detail.
In these cases, the set of equations is equivalent to those obtained in 
the spherical-harmonics expansion.
Stationary and homogeneous solutions are explicitly treated.
In order to solve numerically the equations, a cut in the energy range
is introduced. The modified model maintains the physical characteristics 
of the original equations.
The solution of an asymptotic equation is found and compared to the 
numerical solutions.\\[5mm]
\sez{Introduction}
The motion of electrons in a crystal is governed by complex physical 
laws so that accurate models are required to achieve correct results.
The drift-diffusion model has been widely used in the past, and it has
provided a good description of relevant physical mechanisms.
In modern devices, whose sizes are in the submicron range, 
non-equilibrium effects play an important role and are not adequately 
modeled by the drift-diffusion approach. 
Since in many processes a more accurate description than the 
hydrodynamical setting is required, Boltzmann equation or Monte Carlo 
simulations are employed.
A fully kinetic treatment of carrier dynamics guarantees accurate 
results but requires very expensive numerical procedures in order to 
solve realistic problems. 

To reduce the complexity of the use of the full Boltzmann equation, many 
authors (see ref.~\cita{Ferr} and references therein) have 
introduced simpler models, assuming particular forms for the 
probability density function.
The aim is to yield results, which are more accurate than those 
obtained by hydrodynamical models but less expensive than direct particle 
simulations or numerical treatment of the Boltzmann equation. 
In this framework, a well-known model is derived using a 
spherical-harmonics expansion \cita{SmJe}, \cita{VeBa}, \cita{Fer2}, 
\cita{VeRu} from the Boltzmann equation.

The paper is organized as follows.
In Sec. 2, we introduce the Boltzmann Equation. The collisional operator
takes into account the interactions between 
electrons and phonons, {\it i.e.} vibrations of the lattice. 
This is assumed to be in thermal equilibrium. 
A set of kinetic equations is derived in Sec. 3. 
These are obtained for a general expression of the microscopic kinetic 
energy, so that, for example, they can be used also for non-parabolic bands.
The equations are of partial differential and difference type. 
Equations are specialized in Sec. 4 in the parabolic band 
approximation, and is Sec. 5, where the Kane band model is assumed.
In both cases we show that the equations are equivalent to
those obtained in the spherical harmonics framework.
In Sec. 6 we write the equations in the stationary and homogeneous case. 
The rest of the paper is devoted to the study of these equations.
In Sec. 7 an asymptotic equation, valid for large value of the
electric field, is found and analyzed; 
the explicit solution is found.
In Sec. 8 we discuss some difficulties related to the infinite energy 
range and we propose a suitable solution by introducing a cut in
the scattering rate of the 
collisional operator. 
The numerical scheme to solve the equations is presented in Sec. 9, and 
the results are shown in the last section.

Throughout the paper, boldface and lightface symbols denote vectors and 
scalar quantities respectively. 
\sez{Basic equations} 
We consider an electron gas, which moves in a lattice, subjects to an 
external electric field $\bE$. This can be applied or related, by Poisson 
equation, to the density of the gas.
For the electron gas in a semiconductor device the Boltzmann equation 
\cita{Ferr}, \cita{VeBa} writes
\begin{equation}
 \devp{f}{t} + \frac{1}{\hbar} \nabla_{\bk} \en \cdot \nabla_{\bx} f -
 \frac{\ecar}{\hbar} \bE \cdot \nabla_{\bk} f = Q(f) ,
 \label{e1}
\end{equation}
where the unknown function $\ftxk $ 
represents the probability of finding an electron at the position 
$\bx \in \Rk$, with the wave-vector $\bk \in \Rk$, at time $t$.
The parameters $\hbar$ and $\ecar$ are the Planck constant divided by 
$2\pi$ and the positive electric charge, respectively.
The symbol $\nabla_{\bk}$ stands for the gradient with respect to the 
variables $\bk$ and $\nabla_{\bx}$ that with respect to the 
space coordinates $\bx$. 
The particle energy $\en$ is an assigned nonnegative continuous function, 
defined on $\Rk$, of the wave-vector $\bk$. 
Here we assume that the low-density approximation holds, so that $Q$ is
linear in $f$.
If $\Kkka$ is the sum of the scattering kernels, which describe the 
nature of the inelastic collisions (for example, electron--polar or 
non-polar optical phonon scattering), and $\Kzer$ is the kernel for the 
elastic collisions (for example, electron--impurity scattering), then 
the collisional operator writes \cite{SmJe}, \cite{VeBa}, \cite{JaLu}   
\begin{eqnarray}
& & Q(f)  = (\nq +1) \Ika{\Kkka \ddm  f(t, \bx, \bka) } \\ 
& & \mbox{ } +  
 \nq \Ika{ \Kkak \ddp  f(t, \bx, \bka) } 
\nonumber \\
& & \mbox{ } + \Ika{ \Kzer \dee f(t, \bx, \bka) } - \bar{\nu}(\bk) \ftxk 
\sv \nonumber \label{e3}  
\end{eqnarray}
where
\begin{eqnarray}
& & \mbox{} \hspace{5mm} \bar{\nu}(\bk) = \nq \Ika{\Kkka \ddm}  \\
& & \mbox{} + (\nq + 1) \Ika{\Kkak \ddp} 
+ \Ika{\Kzerin \dee} \p \nonumber  \label{e4}
\end{eqnarray}
Often, in the following we omit to write explicitly that $f$ depends 
also on $t$ and $\bx$, whereas we do not operate with these variables. \\ 
The constant quantity $\nq$ represents the thermal-equilibrium number 
of phonons and is given by
\begin{displaymath}
 \nq = \frac{1}{\exp{ \left( \frac{\hw}{\kT} \right) } - 1 }
\end{displaymath}
where $\kB$ is the Boltzmann constant and $\TL$ is the constant lattice 
temperature.  

The symbol $\dpm$ means the composition of the real-valued function 
$ \enka - \enk \pm \hw $ and the Dirac distribution 
$\delta$. 
Since the function $\en$ may have many different expressions, it 
is not possible in general, by using a unique explicit technique, to 
transform each of the integral operators \eq{e3} in equivalent operators 
without Dirac distributions. 
Therefore, we shall maintain the original form of \eq{e3} in the following.

As usual, due to the isotropy of phase-space, the kernels $\Kkka$ 
and $\Kzer$ are symmetric with respect to the wave-vectors $\bk$ and 
$\bka$.
Moreover, since the kernels are scalar, they depend on $\bk$ and $\bka$ 
only through the scalar quantities $\enk$, $\enka$ and $\bk \cdot \bka$.
This implies in particular that the collision frequency $\nu$ depends on 
$\bk$ only through the variable $\en$. 
\sez{Energy-kinetic equations} 
In the framework of the Boltzmann equation for a perfect gas, a 
classical procedure, due to Grad~\cita{Grad},  
to reduce the dimension of the space of the independent coordinates, 
is the moment method. 
It consists in expanding the ratio between the probability density $f$ 
and a local maxwellian function. Inserting this expansion in the 
Boltzmann equation, an infinite sequence of differential equations are 
obtained. In general no finite set of equations is uncoupled from the 
rest of the system. To obtain a determined system up to certain order
$m$ (usually 13) only $m$ equation are considered  and the coefficients 
of the expansion of higher order are assumed negligible.

In order to obtain a new set of equations, where the unknown functions 
depend on the space-time coordinates and the microscopic kinetic energy, 
we follow the scheme analyzed in \cita{Maj3}. This scheme partially 
recalls the moment method. 

If $\Pk$ is a polynomial and $u$ a real variable, 
we multiply the Boltzmann equation 
\eq{e1} by $\Pk \deu$ and then formally integrate with respect to the
variable $\bk$ over the whole space $\Rk$. 
It is possible to show~\cita{Maj3} that the following equation 
can be derived:
\begin{eqnarray}
& & \devp{\mbox{ }}{t} \Ik{\ftxk \Pk \deu}  \\ 
& & \mbox{} +  \nabla_{\bx} \Ik{\vk \ftxk \Pk \deu} \nonumber  \\
& & - \ecar \bE \cdot \devp{\mbox{ }}{u} 
 \Ik{\ftxk \Pk \vk \deu } \nonumber \\
& & + \frac{\ecar}{\hbar} \bE 
\cdot \Ik{\ftxk \left[ \nabla_{\bk} \Pk \right] \deu} 
  = \Ik{Q(f) \Pk \deu} \p \nonumber \label{e5}
\end{eqnarray}
Here the molecular velocity is defined by means of the formula
\begin{equation}
\vk \ud \frac{1}{\hbar} \nabla_{\bk} \enk \p \label{e2}
\end{equation}
For any choice of the function $\Pk$ a new equation is obtained. As it
happens for the moment method, each equation contains, except for 
particular cases \cita{Maj1},~\cita{Maj2} at least two unknown functions. 
For example, if we choose $\Pk \equiv 1$, then the left hand side of 
eq.~\eq{e5} contains in general the following unknowns 
\begin{eqnarray}
\Ntxu & \ud & \Ik{\ftxk \deu} \sv \label{e6} \\
\Vtxu & \ud & \Ik{\ftxk \vk \deu} \p \label{e7} 
\end{eqnarray}
The scalar quantity $\Ntxu$ is the probability density function to find 
an electron with energy $u$ at time $t$ and position $\bx$. 
The function $N$ is non-negative, because the probability density $f$ 
is always non-negative. 
We note that the presence of the delta-function implies that $N$ and 
$\bV$ vanish for all $\txu$ such that 
$u \leq u_{0} = \min \left\{ \enk : \bk \in \Rk \right\}$.
If the function $N$ is known, then
we can obtain all the hydrodynamical scalars. 
For example, to find the 
hydrodynamical density $\rho \tx$, it is sufficient to integrate $N$ with 
respect to all values of energy $u$; {\it i.e.} 
\begin{displaymath}
\rho \tx \ud \Iu{\Ntxu} \p
\end{displaymath}
The physical meaning of the other quantity $\Vtxu$ can be understood 
with a similar argument.

In the rest of this article we assume the function $\enk$ to be even 
({\it i.e.} $\enk = \en(-\bk)$) and the kernels $K$ and $K_{0}$ to 
depend only on 
$\enk$ and $\enka$. The first hypothesis is verified for the most common 
models used in the numerical simulations. The second 
are introduced in order to simplify the treatment of the equations.
Let 
\begin{displaymath}
\Kb(\enk, \enka) \ud \Kkka \; \mbox{ and } \; \Kb_{0}(\enk, \enka) 
\ud \Kzer \p  
\end{displaymath}
Then eqs.~\eq{e3}-\eq{e4} become
\begin{eqnarray}
 Q(f) & = & 
(\nq +1) \Kb (\enk, \enk + \hw) N(t, \bx, \enk + \hw)   \\
& & \mbox{} + \nq \Kb (\enk - \hw, \enk) N(t, \bx, \enk - \hw) 
\nonumber \\
& & \mbox{} + 
\Kb_{0} (\enk , \enk) N(t, \bx, \enk)  - \nu(\enk) \ftxk
 \nonumber \label{e8}
\end{eqnarray}
with
\begin{eqnarray}
 \nu(\enk) & = & 
(\nq + 1) \Kb (\enk - \hw, \enk) \su{\enk - \hw} \\
& & \mbox{} + \nq \Kb (\enk, \enk + \hw) \su{\enk + \hw} 
+ \Kb_{0} (\enk , \enk) \su{\enk} \sv \nonumber \label{e9}
\end{eqnarray}
where
\begin{equation}
\su{u} \ud \Ik{\deu} \label{e10} 
\end{equation}
is the density of states.\\
\indent
We are interested in deducing two equations in the only variables $N$ and 
$\bV$.
By choosing $\Pk \equiv 1$ and $\Pk = \vk$ eqs.~\eq{e5}, \eq{e8} 
and \eq{e9} give the following equations
\begin{eqnarray}
& & \devp{N}{t} + \nabla_{\bx} \bV - \ecar \bE \cdot \devp{\bV}{u} = 
G(N)  \label{e11} \\
& & \devp{V_{i}}{t} + \sum_{j=1}^{3} \left[ \devp{\Pi_{ij}}{x_{j}} -
\ecar E_{j} \devp{\Pi_{ij}}{u} + \frac{\ecar}{\hbar} E_{j} \Delta_{ij} 
\right] = - \nu(u) V_{i} \quad (i=1,2,3) \sv \label{e12}
\end{eqnarray}
where
\begin{eqnarray}
& &  \mbox{} \hspace{5mm} 
G(N) \ud  (\nq + 1) \Kup \su{u} N(t, \bx, \uph) \\
& &  \mbox{} \hspace{5mm} + \nq \Kum \su{u} N(t, \bx, \umh) 
\nonumber \\
& & \mbox{} \hspace{5mm} - \left[ \nq \Kup \su{\uph} + (\nq + 1) \Kum 
\su{\umh} 
\right] N \txu \nonumber  \sv \label{e13} \\
& &  \mbox{} \hspace{5mm} \nu(u) = \nq \Kup \su{\uph} \\
& & \mbox{} \hspace{5mm} + (\nq + 1) \Kum \su{\umh} + \Kb_{0}(u,u) 
\su{u} \sv \nonumber   \label{e14} \\
& &  \mbox{} \hspace{5mm} \Pi_{ij} \ud \Ik{\ftxk v_{i}(\bk) v_{j}(\bk) 
\deu} \sv \label{e15} \\
& & \mbox{} \hspace{5mm} \Delta_{ij} \ud \Ik{\ftxk 
\devp{v_{i}(\bk)}{k_{j}} \deu} \p \label{e16}
\end{eqnarray}
Also the tensor $\Delta_{ij}$ is symmetric, due to eq.~\eq{e2}.
We note that the effects of the elastic collisions, through the kernel 
$\Kb_{0}$, disappear in the operator $G(N)$. \\
The set of eqs.~\eq{e11}-\eq{e12} contains $N$ and $\bV$ but also two 
tensors  $\Delta_{ij}$ and $\Pi_{ij}$.
Therefore, except in the case of spatially homogeneous solutions with null 
electric field, we need to assume some relations (usually called {\em 
constitutive equations} in the thermodynamic framework), 
which link $\Delta_{ij}$ and $\Pi_{ij}$ with $N$ and $\bV$.
These relations should also depend on the form of the microscopic 
kinetic energy $\en$. 

We feel that at this stage it would be worthwhile to make a simple
observation.
We note that the linearity of the Boltzmann equation \eq{e1} is 
maintained in eqs.~\eq{e11}-\eq{e12}.
To maintain this feature, we suggest to assume constitutive relations of 
the kind
\begin{eqnarray}
\Pi_{ij}\txu & = & p_{ij}(u) N\txu + p_{i}(u) V_{j}\txu + 
  p_{j}(u) V_{i}\txu \label{e17} \\
\Delta_{ij}\txu & = & d_{ij}(u) N\txu + d_{i}(u) V_{j}\txu +
d_{j}(u) V_{i}\txu \sv \label{e18}
\end{eqnarray}
where the symmetric tensors $p_{ij}$ and $d_{ij}$ and the vectors 
$p_{i}$ 
and $d_{i}$ must be determined.
In the next sections we propose a simple choice in the cases of the 
parabolic and non-parabolic band approximation.
\sez{The parabolic case}
The simplest and widely used expression for the microscopic kinetic energy 
\cita{Blak} is 
\begin{equation}
\enk = \frac{\hbar^{2}}{2 \mass} k^{2} \label{par} \sv
\end{equation}
where $\mass$ is the value of the effective electron mass in the 
parabolic mass approximation.
In this case the tensor $\Delta_{ij}$ reduces to
\begin{equation}
 \Delta_{ij} = \frac{\hbar}{\mass} N \txu \delta_{ij}
\quad (i,j = 1, 2, 3) \sv \label{e19}
\end{equation}
where $\delta_{ij}$ is the Kroneker symbol; so that no further 
assumption on eq.~\eq{e18} is required.
The tensor $\Pi_{ij}$ is determined by assuming that \eq{e17} holds with  
$p_{i}(u) \equiv 0$ and moreover that  the constitutive relation {\em is 
exact if $f$ depends on $\bk$ only through the variable $\en$}. 
For $\ftxk = \tilde{f} (t, \bx, \enk )$, we have
\begin{eqnarray*}
N \txu & = & \tilde{f} (t, \bx, u) \su{u} \\
\Pi_{ij} \txu & = & \tilde{f} (t, \bx, u) \Ik{ v_{i}(\bk) v_{j}(\bk) \deu}
\quad (i,j = 1, 2, 3) \sv 
\end{eqnarray*}
so that 
\begin{displaymath}
\su{u} p_{ij}(u) = \Ik{ v_{i}(\bk) v_{j}(\bk) \deu}  \p
\end{displaymath}
Now, it is easy to verify that 
\begin{displaymath}
\su{u} = 4 \sqrt{2} \pi \left( \frac{\sqrt{\mass}}{\hbar} \right)^{3} 
\theta(u) 
\sqrt{u} \sv
\end{displaymath}
where $\theta$ is the Heaviside step function.
Then, for every $u \ge 0$, 
\begin{displaymath}
\Ik{ v_{i}(\bk) v_{j}(\bk) \deu} = \frac{2}{3} \frac{u}{\mass} 
\delta_{ij} \su{u} \quad (i,j = 1, 2, 3) \p 
\end{displaymath}
Hence, eq.~\eq{e12} becomes
\begin{equation}
\devp{\bV}{t} + \frac{1}{3 \mass} \left[ 2 u 
\nabla_{\bx} N - 2 \ecar \bE \devp{\mbox{ } }{u} (u N) + 3 \ecar \bE N 
\right] = - \nu(u) \bV \p \label{e20}
\end{equation}
In the following the functions $N$ and $\bV$ will be the unknowns, which 
must satisfy eqs.~\eq{e11} and \eq{e20}.
It is evident that the knowledge of these quantities cannot provide the 
probability density $f$ uniquely.
Also, it is very hard to prove that, if $N$ and $\bV$, at the initial time, 
derive from a probability density $f$, there exists a non-negative 
probability density, which gives $N$ and $\bV$ for all times. 
\\
Actually the situation is similar to that of the moments in the 
hydrodynamical equations derived by the Grad method.
Here usually, not even for the initial data, the existence of a 
probability density, which gives all the moments occurring in the 
equations, is considered.

Therefore, we limit to assuming only the simple conditions that $N$ is 
nonnegative and both the variables $N$ and $\bV$ vanish as $u \leq 0$.

We note that these unknowns are simply related to the first two terms 
of the spherical  harmonic expansion of the 
distribution function 
\begin{equation}
 \ftxk = f_{0}(t, \bx, \enk) + \bk \cdot {\bf f}_{1}(t, \bx, \enk) 
 + ...  \p
\label{she}
\end{equation}
It is a simple matter to show that
\begin{eqnarray*}
f_{0}(t, \bx, \enk) & = & \frac{N(t, \bx, \enk)}{\su{\enk}} \sv \\
{\bf f}_{1}(t, \bx, \enk) & = & \frac{3 \hbar }{2 \enk \su{\enk}} 
\bV(t, \bx, \enk) \p
\end{eqnarray*}
The corresponding equations for $f_{0}$ and ${\bf f}_{1}$ 
(see ref.~\cita{RaWA}) 
coincide with our eqs.~\eq{e11} and \eq{e20}.
This is a consequence of the choice on the approximation of the tensor 
$\Pi_{ij}$.
It is evident  that different choices give different sets of equations.
In this paper we do not investigate this possibility. 

Similar equations were proposed by H\"{a}nsch in ref.~\cita{Hans} in the 
framework of quantum kinetic transport. Also in that case a problem 
of equating the number of equations and unknowns
arose. It was solved by simple physical arguments, but the 
equations are different from eqs.~\eq{e11}-\eq{e20}. 
\sez{The non-parabolic band approximation}
For large values of electron energy eq.~\eq{par} is not adequate. 
The following suitable expression was given by Kane 
\cita{Fer2},~ \cita{Blak}: 
\begin{equation}
\en(1+\alk \en) = \frac{\hbar^{2}}{2 \mass} k^{2} \label{nonpar} \sv
\end{equation}
where $\alk$ is a constant. Now, it is simple to verify that 
\begin{displaymath}
\su{u} = 
4 \sqrt{2} \pi \left( \frac{\sqrt{\mass}}{\hbar} \right)^{3} 
\theta(u) 
\sqrt{u(1+\alk u)}(1 + 2 \alk u) \p
\end{displaymath}
So eq.~\eq{e11} can be written explicitly. The second equation 
requires only to determine $\Pi_{ij}$ and  $\Delta_{ij}$ 
according to eqs.~\eq{e17}-\eq{e18}. Due to the spherical
symmetry of the band we assume $p_{j}(u)=d_{j}(u)=0$.
Analogously to the parabolic case, we require that 
eqs.~\eq{e17}-\eq{e18} hold if $f$ depends on $\bk$ only
through $\en$. In the same manner we can see that that
\begin{displaymath}
p_{ij}(u) = g(u)\delta_{ij} ~~~,~~~ d_{ij}(u) = h(u)\delta_{ij}
\end{displaymath}
where
\begin{displaymath}
g(u) = \frac{2}{3}~\frac{u(1+\alk u)}{\mass(1+2 \alk u)^{2}} 
\end{displaymath}
and
\begin{displaymath}
h(u) = \frac{1}{\mass(1+2 \alk u)} 
- \frac{4 \alk }{3 \mass}~ 
\frac{u(1+\alk u)}{(1+2 \alk u)^{3}} \p
\end{displaymath}
Therefore eq.~\eq{e12} becomes
\begin{equation}
\devp{\bV}{t} +  g(u) 
\nabla_{\bx} N - \ecar \bE \devp{\mbox{ } }{u} 
\left[ g(u) N \right] +  \ecar \bE~ h(u) N 
 = - \nu(u) \bV \p \label{e20np}
\end{equation}
Also in this case the equations for $N$ and $V$ are equivalent to
those obtained by means of the spherical harmonics expansion
\cita{RaWA}.
For $\alk=0$ we recover eq.~\eq{e20}.
\sez{Homogeneous solutions with constant electric field}
We consider equations \eq{e11} and \eq{e20np}, and 
we look for a solution depending only on $u$. 
Since in this case the electric field must be a constant vector, we 
choose the reference frame so that only the first component of $\bE$ is 
different from zero. As a consequence the only significant component of 
$\bV$ is the first.
We limit to assume constant kernels $\Kb$ and $\Kb_{0}$. This fact
holds when a homogeneous, intrinsic silicon at room temperature is
considered.

It is useful to introduce dimensionless variables.
Let be
\begin{displaymath}
 t_{*} \ud \left[ 4 \sqrt{2} \pi 
 \left( \frac{\sqrt{ \mass} }{\hbar} \right)^{3} \sqrt{\kT} \Kb \nq 
\right]^{-1} \sv \quad
\ell_{*} \ud \sqrt{\frac{\kT}{\mass}} t_{*} \sv \quad
u_{*} \ud \kT \sv
\end{displaymath}
\begin{displaymath}
w \ud \frac{u}{u_{*}} \sv \quad
n(w) \ud u_{*} \ell_{*}^{3} N(u) \sv \quad
v(w) \ud u_{*} \ell_{*}^{2} t_{*} V_{1}(u) \sv \quad
\end{displaymath}
\begin{displaymath}
\alpha \ud \frac{\hw}{\kT} \sv \quad
a \ud \frac{\nq + 1 }{\nq} = e^{\alpha} \sv \quad
\beta = \frac{\Kb_{0}}{\nq \Kb} \sv \quad 
\alt \ud \alk u_{*} \sv \quad
\zeta \ud \ecar E \frac{\ell_{*}}{u_{*}}  \p
\end{displaymath}

It is a simple matter that the following equations can be obtained
\begin{eqnarray}
& \hspace{-20pt} & - \zeta \devo{v}{w} = s(w) \left[ a n(\wpa) 
+ n(\wma) \right] 
- \left[ s(\wpa) + 
a s\left(\wma \right) \right] n(w) \label{e21} \\
& \hspace{-20pt} & \zeta \left[ - \devo{\mbox{ }}{w}(r(w) \, n) 
+ q(w) n \right] = - \left[ s(\wpa) + a s\left(\wma \right) + \beta s(w)
\right] v(w) \label{e22} \sv
\end{eqnarray}
where
\begin{eqnarray}
s(w) &=& 
\theta(w) 
\sqrt{w(1+\alt w)}(1 + 2 \alt w) \\
r(w) &=& \frac{2}{3}~\frac{w(1+\alt w)}{(1+2 \alt w)^{2}} \\
q(w) &=& \frac{1}{1+2 \alt w} 
- \frac{4 \alt }{3}~ 
\frac{w(1+\alt w)}{(1+2 \alt w)^{3}} \p
\end{eqnarray}
As previously mentioned, we recall that we assume the conditions 
\begin{displaymath}
n(0) = v(0) = 0  \p
\end{displaymath}
We note that in this case eq.~\eq{e22} gives $v(0) = 0$ as a consequence
of the boundary condition $n(0)=0$.

We look for solutions of eqs.~\eq{e21}-\eq{e22} satisfying the following 
conditions
\begin{eqnarray}
& & n(0) = 0 \sv \qquad \lim_{w \freccia + \infty} v(w) = 0 \label{e23}
\\
& & n(w) \geq 0 \perogni w \geq 0 \mbox{ and } \int_{\rind} n(w) \, dw > 
0 \p \label{e24}
\end{eqnarray}
Since eqs.~\eq{e21}-\eq{e22} are linear and homogeneous, if a solution 
$(n(w), v(w))$, satisfying the above conditions exists, then, for every 
$c > 0$,  also $(c \hspace{1pt} n(w), c \hspace{1pt} v(w))$ is 
solution. 

It is important to note that {\em the conservation of particle number} holds 
for the system \eq{e21}-\eq{e22}. 
In fact, from eq.~\eq{e21} it is easy to verify that
\begin{equation}
\Iw{ s(w) \left[ a n(\wpa) 
+ n(\wma) \right] - \left( s(\wpa) + 
a s(\wma) \right) n(w) } = 0 \sv  \label{e25}
\end{equation}
for any nonnegative function $n$ such that $n(w)s(w)$ is integrable  
and $n(0)=0$.
This property forces to choose the boundary on $v$ in \eq{e23}, being 
$v(0) = 0$.
\sez{Asymptotic equations}
A simple approximate solution of the previous problem can be
obtained in the parabolic band case.
We introduce a new variable $\psi$ defined by
\begin{displaymath}
n(w) = \sqrt{w} \psi(w) \p
\end{displaymath}
Then, taking into account that now $\alt=0$,
eqs.~\eq{e21}-\eq{e22} become
\begin{eqnarray}
& \hspace{-20pt} & - \zeta \devo{v}{w} = \sqrt{w} \left[ a \sqrt{
\wpa} \psi(\wpa) 
+ \sqrt{\wmap} \psi(\wmap) \right]  \\ 
& & \hspace{30pt} \mbox{ } - \sqrt{w} \left( \sqrt{\wpa} + 
a \sqrt{\wmap} \right) \psi(w) \nonumber \label{e26} \\
& \hspace{-10pt} & \frac{2}{3} \zeta \sqrt{w^{3}} \, \devo{\psi}{w} = 
\left( \sqrt{\wpa} + a \sqrt{\wmap} + \beta \sqrt{w} 
\right) v(w) \label{e27} \sv
\end{eqnarray}
where $(z)_{+} = \max\{z,0\}$.\\
We look for asymptotic equations of system ~\eq{e26}-\eq{e27} for large 
values of the energy $w$.
To this scope we expand the coefficients of the equations up to the zeroth 
order (for example $\sqrt{\wpa} \simeq \sqrt{w}$).
We obtain a new set of equations
\begin{eqnarray}
 - \zeta \devo{v_{A}}{w} & = & w \left\{ \left[ a  \psi_{A}(\wpa) 
+ \psi_{A}(w - \alpha) \right]  - (a+1) \psi_{A}(w) \right\} \label{e28} \\
 v_{A}(w) & = &  \frac{2 \zeta}{3 (1 + a + \beta)} w 
\devo{\psi_{A}}{w}  \label{e29} \sv
\end{eqnarray}
where the subscript $A$ indicates the new unknowns.
Inserting $v_{A}(w)$ in the first equation and neglecting a coefficient 
proportional to $w^{-1}$, a single equation in $\psi_{A}$ is obtained
\begin{equation}
\frac{2 \zeta^{2}}{3(a + 1 + \beta)} \frac{d^{2} \psi_{A}}{d \, w^{2}} + 
a \psi_{A}(\wpa) + \psi_{A}(w - \alpha) - (a+1) \psi_{A}(w) = 0 
\p \label{e30}
\end{equation}
This is a linear difference-differential equation. It is easy to see 
that $\exp(\lambda w)$ is a solution of eq.~\eq{e30} if and only if 
$\lambda$ satisfies the transcendent equation
\begin{equation}
\frac{2 \zeta^{2}}{3(a + 1 + \beta)} \lambda^{2} +  a e^{\lambda \alpha} 
+ e^{ - \lambda \alpha} - (a+1) = 0 \p \label{e31}
\end{equation}
In Appendix A, we prove that this equation admits only two solutions: 
$\lambda = 0$ and $\bar{\lambda} \in (-1,0)$. 
We give also a simple approximation for $\bar{\lambda}$.
\\
By means of this solution of eq.~\eq{e30}, we get the original quantities 
\begin{equation}
n_{A}(w) = c \sqrt{w} e^{\bar{\lambda} w} \sv \quad
v_{A}(w) = \frac{2 \zeta w}{3 (1 + a + \beta)} \bar{\lambda} c 
e^{\bar{\lambda} w} \sv \label{e32}
\end{equation}
where $c$ is a positive constant, in order to satisfy 
the condition \eq{e24}.
Despite this solution is obtained for large values of $w$ it often 
agrees with the numerical results (which we present in next section) 
in the whole range of $w$.
\\
In the limit case $\zeta = 0$ (no electric field) $\bar{\lambda} = -1$ 
and the solution of \eq{e30} gives the correct function $N$ obtained 
from the equilibrium maxwellian distribution function~\cita{Maj2}.

For high electric fields approximate solutions are known
(see \cite{SmJe}, \cite{Fer2} for example). These are obtained by means 
of an asymptotic expansion of the original differential-difference 
equations, which  are transformed in simple ordinary differential 
equations and are explicitly  solved. 
Our solution, obtained by the asymptotic equation, is different with 
respect to the previous ones, because we do not approximate the 
difference terms $f_{0}(\epsilon \pm \hbar\omega)$ ($\omega$ is the 
constant optical phonon frequency), appearing in the equations, using 
Taylor formula. In fact the asymptotic equation remains of 
differential-difference type.

We are interested in calculating the hydrodynamical velocity of electrons
\begin{displaymath}
 v_{h} \ud \frac{\dm \Iu{V_{1}(u)}}{\dm \Iu{N(u)}} = 
\sqrt{\frac{\kT}{\mass}} \frac{\dm \Iw{v(w)}}{\dm \Iw{n(w)}} \p
\end{displaymath}
By a simple calculation, it is possible to verify that, using \eq{e32},
the following value is obtained
\begin{displaymath}
\left| v_{A_{h}} \right| =  \sqrt{\frac{\kT}{\mass}}
 \frac{4}{3 (1 + a + \beta) \sqrt{\pi}} \zeta \sqrt{| \bar{\lambda} |} \p
\end{displaymath}
If we use the approximate value (see Appendix A) of $\bar{\lambda}$, we 
obtain the saturation velocity
\begin{displaymath}
\lim_{\zeta \freccia + \infty} \left| v_{A_{h}} \right| =
  \sqrt{\frac{\kT}{\mass}} 
\sqrt{\frac{8 \alpha (a - 1)}{3 \pi (1 + a + \beta)}} \p
\end{displaymath}
\sez{Finite energy model}
The difficulties to solve numerically eqs.~\eq{e21}-\eq{e22} are given by 
the deviating arguments in the right side of equations and by the 
infinite interval for the variable $w$.
The first suggests to use a difference scheme to discretize the 
equations with the constraint that the step $\Dw$ be such that $\dm 
\frac{\alpha}{\Dw}$ is an integer.
The second difficulty is solved by considering a finite interval for the 
variable $w$. This truncation must be made carefully. 
To explain this aspect of the problem, we recall that we are looking for 
a solution which describes the no-runaway phenomena. 
It is a particular configuration of the gas, where the applied electric 
field is balanced by the effect of the collisions with the background. 
In other words, the energy that an electron receives from the external 
electric field is transferred to the phonons so that a 
spatial homogeneous and static configuration is maintained.

In order to find this solution also in the case of a finite range for 
the energy $w$, we introduce a cut in the kernels $\Kb$ and $\Kb_{0}$.
Let us choose a positive integer $\Nse$.
Put $M = \Nse \hw$, we consider the interval $[0, M]$ and the 
function $H(u)$ which coincides with the Heaviside step function 
$\theta(u)$ except for $u=0$ where $H(u) = 0$.
Then the new kernels are defined by
\begin{eqnarray*}
\Kb^{M}(\enk, \enka) & = & \Kb(\enk, \enka)
 H ( M - \max \{ \enk, \enka \} ) \sv
\\
\Kb_{0}^{M}(\enk, \enka) & = & \Kb_{0}(\enk, \enka)
H ( M - \max \{ \enk, \enka \} ) \p
\end{eqnarray*}
These kernels imply that a particle has zero probability of collision in 
the following two cases:
\begin{itemize}
\item
the particle before the collision has an energy less than $M$, but after 
it should have an energy equal or greater than $M$;
\item
the particle before the collision has an energy equal or greater 
than $M$.
\end{itemize}
Using the new kernels eqs.~\eq{e21}-\eq{e22} are substituted by the 
following
\begin{eqnarray}
& \hspace{-20pt} & - \zeta \devo{v}{w} = A(n)(w) \label{e33} \\
& \hspace{-20pt} & \zeta \left[ - \devo{\mbox{ }}{w}(r(w) \, n) 
+ q(w)n \right] = - B(v)(w) \label{e34} \sv
\end{eqnarray}
where
\\
$ \dm A(n)(w) \ud s(w) \left[ a H(\alpha \Nse - w - \alpha) 
n(\wpa) + H(\aN - w) n(\wma) \right]$
\\[2mm]
 $ \dm \mbox{} \hspace{22mm} - 
\left[ H(\aN - w - \alpha) s(\wpa) +  a H(\aN - w) s(\wma) 
\right]  n(w)$ \\[2mm]
$ \dm B(v)(w) \ud \left\{ H(\aN - w - \alpha) s(\wpa)  + 
  H(\aN - w) \left[ a s(\wma) + \beta s(w) \right] 
\right\} v(w) \p  $
\\
The corresponding boundary conditions becomes
\begin{eqnarray}
& & n(0) = 0 \sv \qquad  v(\aN) = 0 
\\
& & n(w) \geq 0 \perogni w \geq 0 \mbox{ and } 
\int_{0}^{\aN} n(w) \, dw > 0 \p 
\end{eqnarray}
The boundary condition on $v$ is the natural consequence of the condition 
\eq{e23}.
The choice of the above boundary conditions and the cut are suggested by a 
simple physical picture.
No-runaway phenomena may be considered as the limit for $t \freccia + 
\infty$ of a suitable spatial homogeneous solution of 
eqs.~\eq{e11},~\eq{e20}.
For time-depending solutions, the new kernels and the condition $v(\aN) = 
0$ imply that a particle with initial energy less than $M$ will have, for 
all time, energy less than $M$. For the same reason particles having 
energy greater than $M$ will have, for all time, energy greater than 
$M$. Then, the total number of the particles having energy less than $M$ 
will be constant in time. Different choices can violate the conservation
of the particle number for electrons having energy less than $M$. 
In these cases the number of particles having energy less than $M$ could 
tend to zero or to infinity.
\sez{Numerical solutions}
In order to discretize eqs.~\eq{e33}-\eq{e34}, we fix the step $\Dw$ 
and consider in the interval $[0, \aN]$ the points
\begin{displaymath}
w_{i} = i \times \Dw  \qquad (i=0, 1, 2, ...., \np)
\end{displaymath}
where $\dm \np = \frac{\aN}{\Dw}$.
If we integrate eqs.~\eq{e33}-\eq{e34} in the interval $[ w_{i} , w_{i+1} ]$
exactly where it is possible and using trapezoidal rule otherwise, we 
obtain the linear algebraic system
\begin{eqnarray}
& & \hspace{-20pt}
\frac{2 \zeta}{\Dw} (v_{i+1} - v_{i} ) + A_{i+1} + A_{i} = 0 
\quad (i = 0, 1, ..., \np -1) \sv \label{e35} \\
& & \hspace{-20pt} v_{0} = 0 \label{e36} \\
& & \hspace{-20pt}
\frac{2 \zeta}{\Dw} ( n_{i+1} r_{i+1} - n_{i} r_{i} ) - 
\zeta (q_{i+1} n_{i+1} + q_{i} n_{i}) 
- B_{i+1} - B_{i}= 0 \\ 
& & \hspace{-20pt}(i=1, 2, ..., \np - 1) \sv \nonumber  \label{e37}
\end{eqnarray}
where, for any function $\varphi (w)$, $\varphi_{i} = \varphi(w_{i})$.
The corresponding boundary conditions are
\begin{eqnarray}
& & n_{0} = 0 \sv \quad v_{\nps} = 0 \label{e38} \\
& & n_{i} \geq 0 \quad i=0, 1, ... \np \quad \mbox{and }
\frac{\Dw}{2} \sum_{i=0}^{\nps - 1} \left( n_{i+1} + n_{i} \right) > 0
 \p \label{e39}
\end{eqnarray}
Equations \eq{e37} are obtained considering only the interval 
$[w_{i}, w_{i+1}]$ for $i \geq 1$, because in the first point $w_{0}$ we 
have written the exact equation $v_{0} = 0$ ({\it i.e.} eq.~\eq{e36}), 
which derives from eq.~\eq{e34} with  the boundary condition $n(0) = 0$.
It is simple to deduce from \eq{e35} the following equations
\begin{equation}
v_{j} = v_{\nps} + \frac{\Dw}{2 \zeta} \sum_{i=j}^{\nps -1} (A_{i} + 
A_{i+1} ) \quad (j = 0, 1, ..., \np -1)  \label{e40} \p
\end{equation}
Moreover, as a consequence of the appropriate choice of the cut, the 
conservation of the particle number holds (see ref.~\cita{Maj1}), being
\begin{equation}
\sum_{i=0}^{\nps -1} (A_{i} + A_{i+1} ) = 0 \label{e41} \p
\end{equation}
Eq.~\eq{e41} is the corresponding of eq.~\eq{e25} introducing the 
cut in the kernels and performing the integral using the trapezoidal 
rule.
\\
Now, eq.~\eq{e40} gives $v_{0} = v_{\nps}$ for $j = 0$.
Due to the boundary condition $v_{\nps} = 0$, there follows that 
\eq{e36} is a consequence of eqs.~\eq{e40}, or equivalently eq.~\eq{e40} 
for $j=0$ coincides with eq.~\eq{e36}. For numerical convenience, we 
prefer to eliminate the case $j=0$ in \eq{e40}. 
Now, we must add the condition \eq{e39}.
To select an unique solution, we choose the equation $n_{1} = 1$.
This, with the conditions $n_{i} \geq 0$ for $i=2, 3,...   \np$ 
guarantees \eq{e39}.
This makes the number of equations equal to the number of the unknowns. 
For the sake of clarity, we rewrite the complete linear system
\begin{eqnarray}
& & \hspace{-15pt}
v_{\nps} = 0 \sv \quad v_{0} = 0 \sv \quad n_{0} = 0 
\sv \quad n_{1} = 1 \sv \\
& & \hspace{-15pt}
v_{j} = v_{\nps} + \frac{\Dw}{2 \zeta} \sum_{i=j}^{\nps -1} (A_{i} + 
A_{i+1} ) \quad ( j = 1, 2, ..., \np -1 )  \\
& & \hspace{-15pt}
 \frac{2 \zeta}{\Dw} \left[ n_{i+1} r_{i+1}  - n_{i} 
r_{i}\right] - \zeta (q_{i+1} n_{i+1} + q_{i }n_{i}) 
- B_{i+1} - B_{i}= 0  \\
& & \hspace{-15pt}( i=1, 2, ..., \np - 1 ) \p \nonumber    \label{e42}
\end{eqnarray}
\vspace{-40pt} \mbox{}
\sez{Numerical results and conclusion}
We are interested into solve eqs.~\eq{e33}-\eq{e34} 
in the case of a silicon bulk device. The appropriate values
for the parameters are given in the following table.
\begin{center}
\begin{tabular}{|l|l|l|}
\hline
$ \mass = 0.32 \, m_{e}$ & $ \TL = 300 $ K & $\hw = 0.063$ eV \\[7pt]
$ \dm \Kb = \frac{\left( D_{t} K \right)^{2}}{8 \pi^{2} \rho \omega} $ &
$ D_{t} K = 11.4 $ eV \mbox{$\stackrel{\circ}{\mbox{\rm A}}$}$^{-1}$ &
$\rho = 2330$ Kg m$^{-3}$ \\[15pt]
$ \dm \Kb_{0} = \frac{\kT}{4 \pi^{2} \hbar v_{0}^{2} \rho} \Xi_{d}^{2} $ 
& $\Xi_{d} = 9$ eV & $v_{0} = 9040$ m sec$^{-1}$. \\[15pt]
$ \alk = 0.5 \, eV^{-1}$ & $ \mbox{ } $ & $ \mbox{ } $  \\[7pt]
\hline
\end{tabular}
\end{center}
Here, $m_{e}$ denotes the electron rest mass.
Using these parameters, we get $\alpha \simeq 2.437$ and $\beta \simeq 
5.986$.
The values for the electric field are 
$ | \bE | = 10^{3}$, $10^{4}$ and $10^{5}$ ~Vcm$^{-1}$.
Except for low electric field there is a great difference between P 
(parabolic) and NP (non-parabolic) band.
In fig.~1 we plot the velocity versus the electric field,
obtained using the asymptotic solutions \eq{e32} and values yielded by
numerical integration of eqs.~\eq{e33}-\eq{e34} both in P and NP cases. 
In the first case the value for the saturation velocity obtained is 
$1.27 \times 10^{5} ~m/sec$. 
The second model does not show a limit value as in the
Monte Carlo simulations \cita{Tom}. For the largest value of $ | \bE | $ the
velocity is $0.95 \times 10^{5} ~m/sec$
The experimental value is $1 \times 10^{5 } ~m/sec$.
In order to compare numerical and asymptotic solutions, we have 
chosen the condition
\begin{equation}
\int_{0}^{\aN} n_{A}(w) \, dw = \int_{0}^{\aN} n(w) \, dw
\label{e43}
\end{equation}
where $n_{A}(w)$ indicates the asymptotic solution.
The integral in \eq{e43} of the function $n_{A}$ is performed 
analytically and the second numerically. \\
In such way the constant $c$ in~\eq{e32} is chosen.
The meaning of eq.~\eq{e43} is evident and coherent with 
the criterion of the choice of the cut.
We have made many numerical experiments, varying the values 
of the electric field, the step $\Delta w$ and the cut $\Nse$. 
In fig.~2 the quantity $N(u)$, both for P and NP approximation, is shown 
for a low value of the electric field.
 Nevertheless a small range of the energy 
$ 0 - 4\hw $ is sufficient to contain almost all the electrons,
the numerical results for P case agree with the asymptotic solution.
This is due to the fact that for a small values of the electric
field $\bar{\lambda} \freccia -1$.
The quantity $V_{1}(u)$ is shown in fig.~3.
Here the asymptotic solution is given by the formula
\begin{displaymath}
v_{A}(w) = \frac{2 \zeta \sqrt{w^{3}} 
c \bar{\lambda} e^{\bar{\lambda} w}}
{3 \left( \sqrt{\wpa} + a \sqrt{\wma} + \beta \sqrt{w} \right) }
\end{displaymath}
This choice is not coherent with the expansion, but it allows a better
agreement with the numerical solutions of P case and it puts in evidence 
a discontinuity in the derivative at the point $ w = \alpha  
\iff u = \hw $, due to the term $ \sqrt{\wma} $.  
This square root is also present in the complete set of equations 
\eq{e21}-\eq{e22}. Then, such discontinuity is also expected
in the solutions.
A deeper analysis could be necessary to prove mathematically 
this fact.
Figs. 4 and 5 shown an intermediate case where the agreement is 
poor. \\
When the electric field is high (figs. 6 and 7) very small
variations are observable. This agreement was reasonable for 
large values of $u$.
In the other cases this is due to the number of particles having
low energy which is enough small with respect to that of the whole
gas.\\
The last figures show a quite surprising result in the parabolic case. 
We have used the
same value of the electric field $E$ as in fig. 6 and 7. 
Here, a small value of $\Nse$ is used. 
The density $N$ agrees with the asymptotic solutions. The
quantity $V_{1}$ subject to the boundary condition 
$V_{1}(\Nse \alpha) = 0 $, 
follows the asymptotic solutions almost to the end of the 
interval. \\[25pt]
{\large \bf Acknowledgments}
\\
We acknowledge partial support from Italian Consiglio Nazionale
delle Ricerche (Prog. N. 96.03855.CT01) and TMR project 
n.~ERBFMRCT970157 ``Asymptotic Methods in Kinetic Theory''.\\[10pt]
{\large \bf Appendix A} 
\\
The function
\begin{displaymath}
\eta(\lambda) = \frac{2 \zeta^{2}}{3(a + 1 + \beta)} \lambda^{2} 
+  a e^{\lambda \alpha} 
+ e^{ - \lambda \alpha} - (a+1)  
\end{displaymath}
for $\zeta \geq 0$ is convex, because $ \eta''(\lambda) \geq 0$
for every real $\lambda$. \\
Since $\eta(0)=0$ and $\eta'(0) =\alpha(a-1) > 0$, then there follows 
that the minimum is negative and is reached at 
a point $\lambda_{m} < 0$.\\  
From $\eta(-1) = 2 \zeta^{2} \left(3(a + 1 + \beta) \right)^{-1}$ 
(recall that $ a = e^{\alpha} $) we get $-1 < \lambda_{m} <0$.   
Then, there exists only  $\bar{\lambda} \ne 0$ such that 
$\eta(\bar{\lambda}) = 0$. Moreover $\bar{\lambda} \in [-1,0[$.
An approximation of this value is obtained expanding 
$e^{\pm \lambda \alpha}$ around $\lambda = 0$ up the 
term of second order. A second order algebraic equations follows 
\begin{displaymath}
\left[\frac{2 \zeta^{2}}{3(a + 1 + \beta)} +
\frac{\alpha^{2}}{2}(a+1)\right] \lambda^{2} +
\alpha(a-1)\lambda = 0  \p
\end{displaymath}
Then
\begin{displaymath}
\bar{\lambda} \simeq -\frac{6 (a + 1 + \beta) \alpha (a-1)}
{4 \zeta^{2} + 3 (a + 1 + \beta) \alpha^{2} (a+1) }  
 \p
\end{displaymath}
For $\zeta \gg 1$ we get
\begin{displaymath}
\bar{\lambda} \simeq -\frac{3\alpha(a-1)(a+1+\beta)}{2\zeta^{2}}
\p
\end{displaymath}
%
%
%
%
%

%
\newpage
\begin{figure}[h!]
\centerline{\psfig{figure=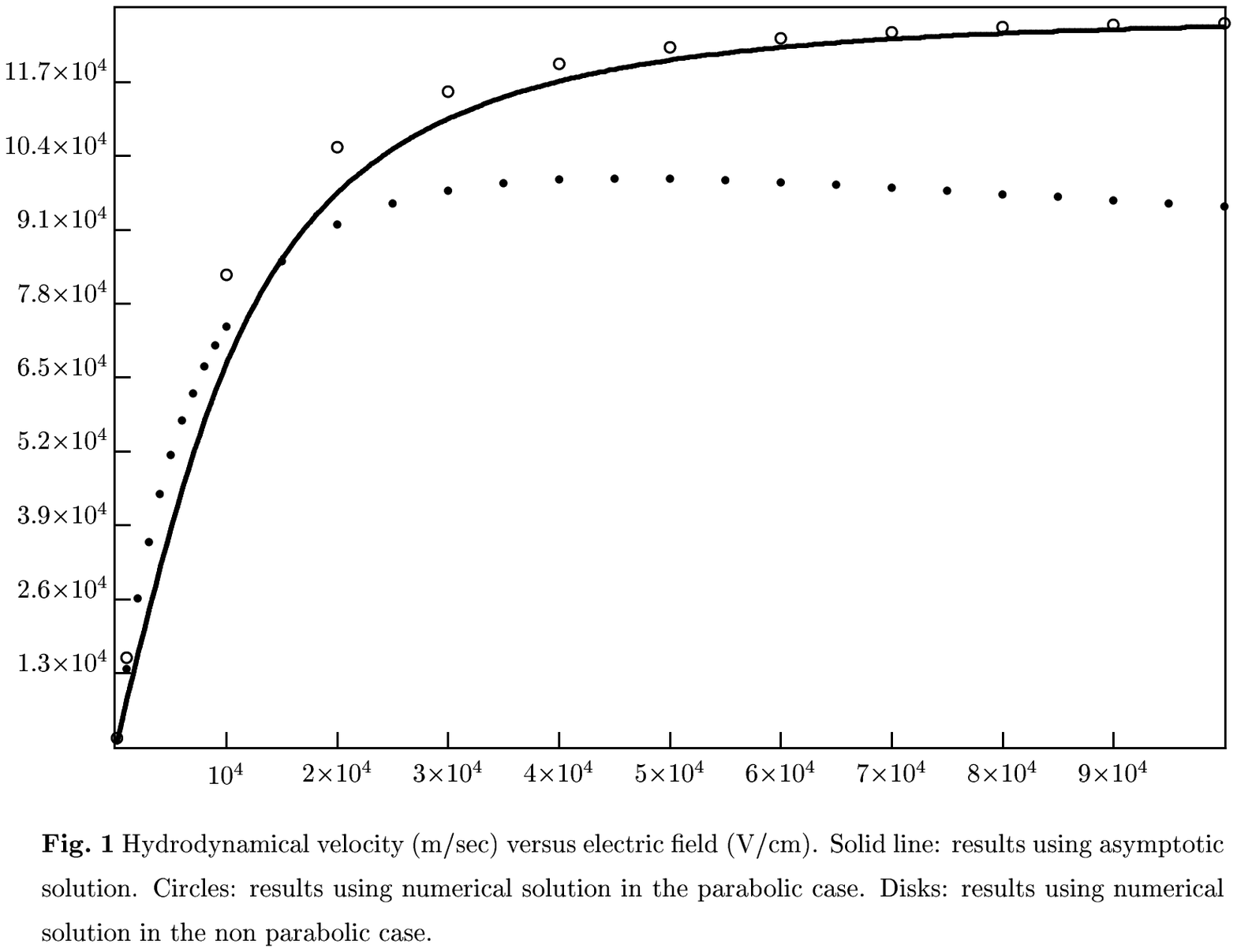}}
\end{figure}
\newpage
\begin{figure}[h!]
\vspace{-5.2cm}
\centerline{\psfig{figure=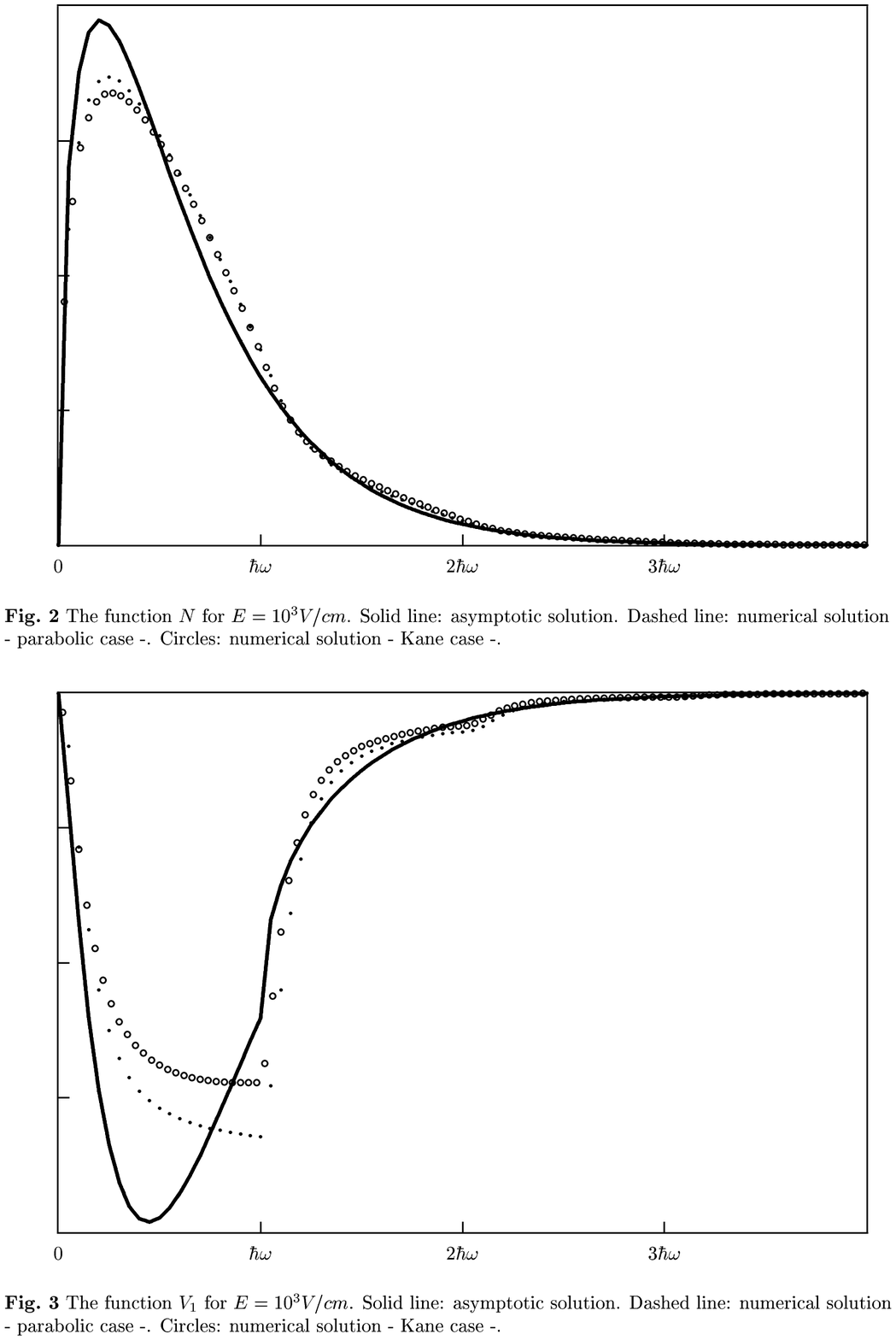}}
\end{figure}
\newpage
\begin{figure}[h!]
\vspace{-5.2cm}
\centerline{\psfig{figure=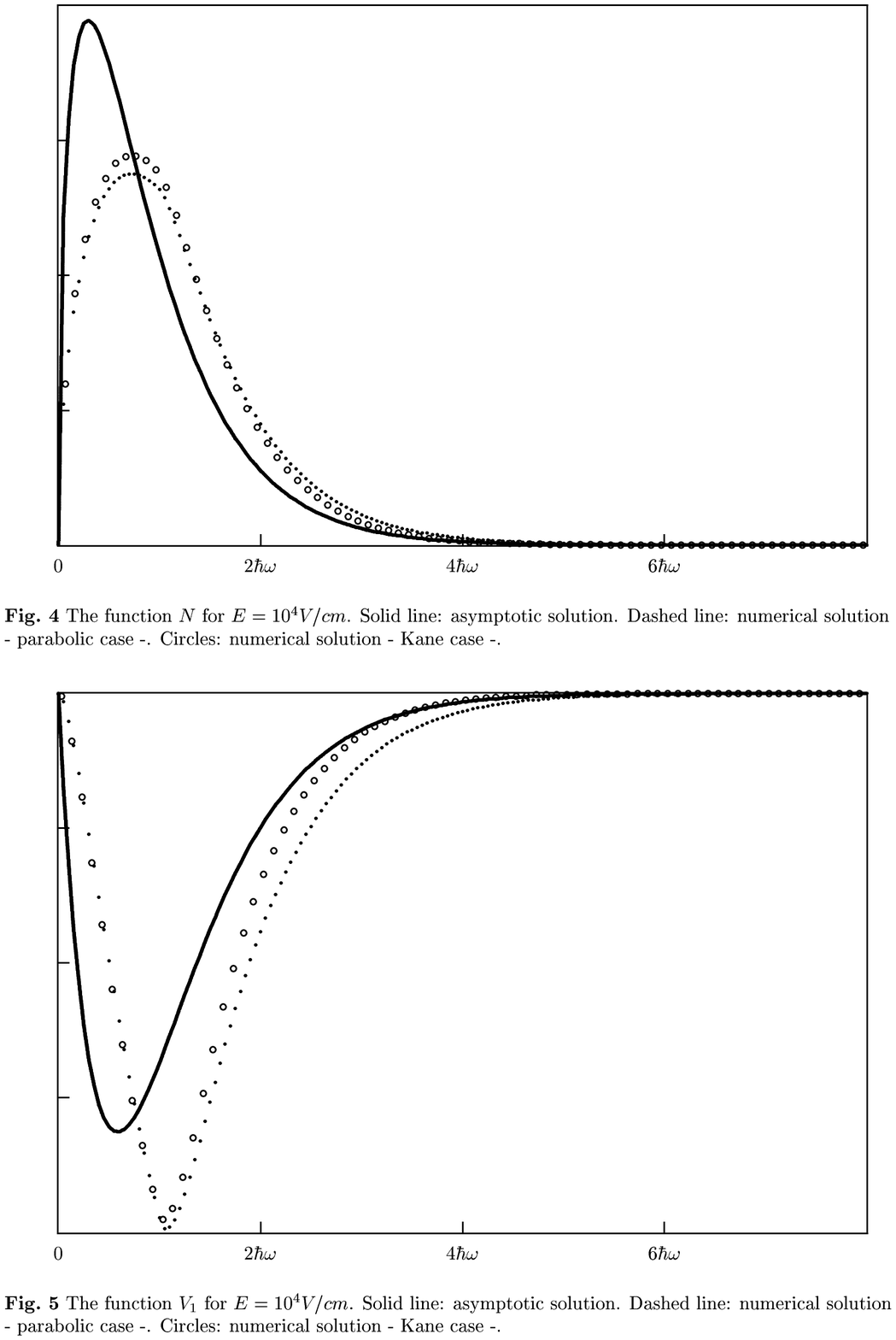}}
\end{figure}
\newpage
\begin{figure}[h!]
\vspace{-5.2cm}
\centerline{\psfig{figure=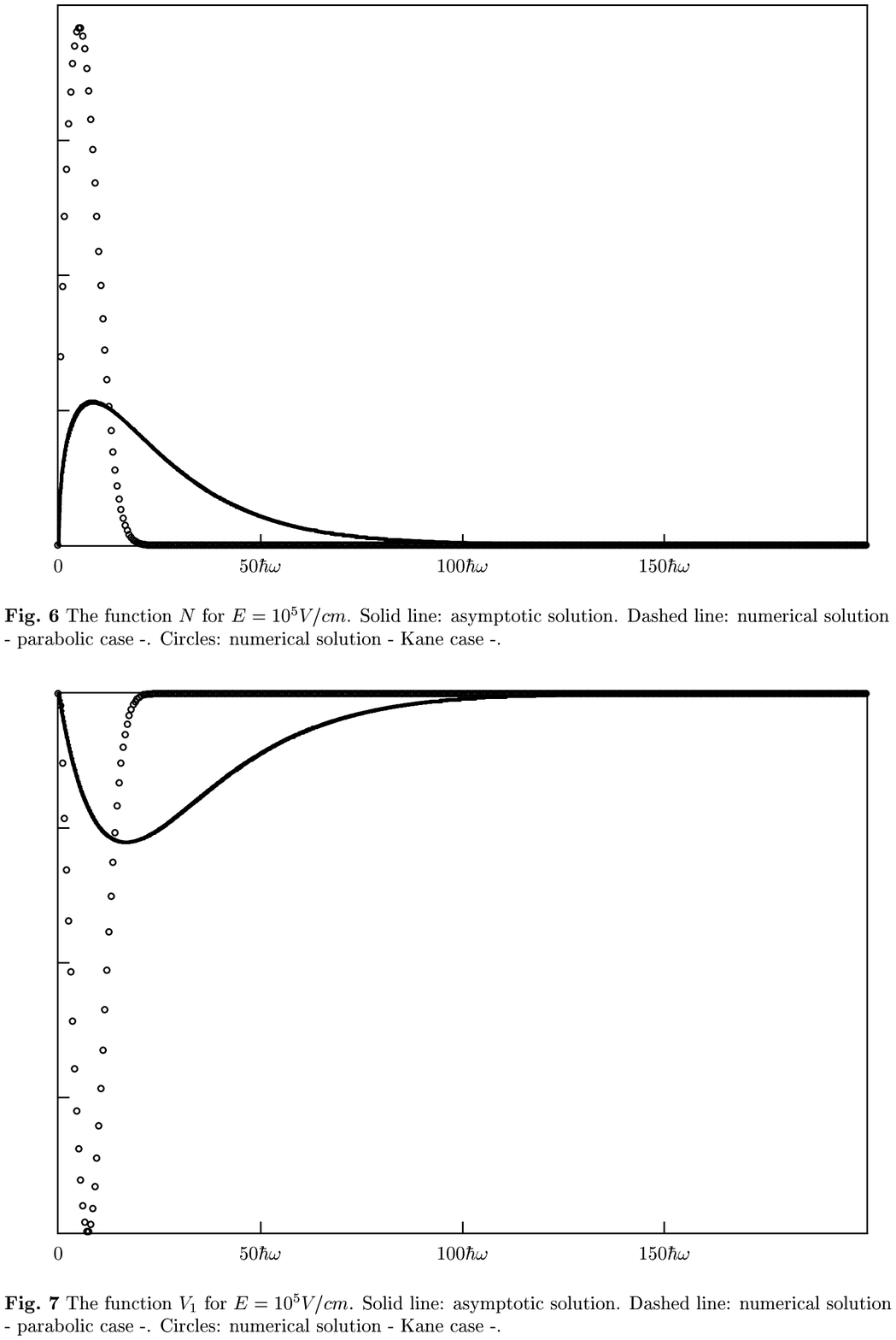}}
\end{figure}
\newpage
\begin{figure}[h!]
\vspace{-5.2cm}
\centerline{\psfig{figure=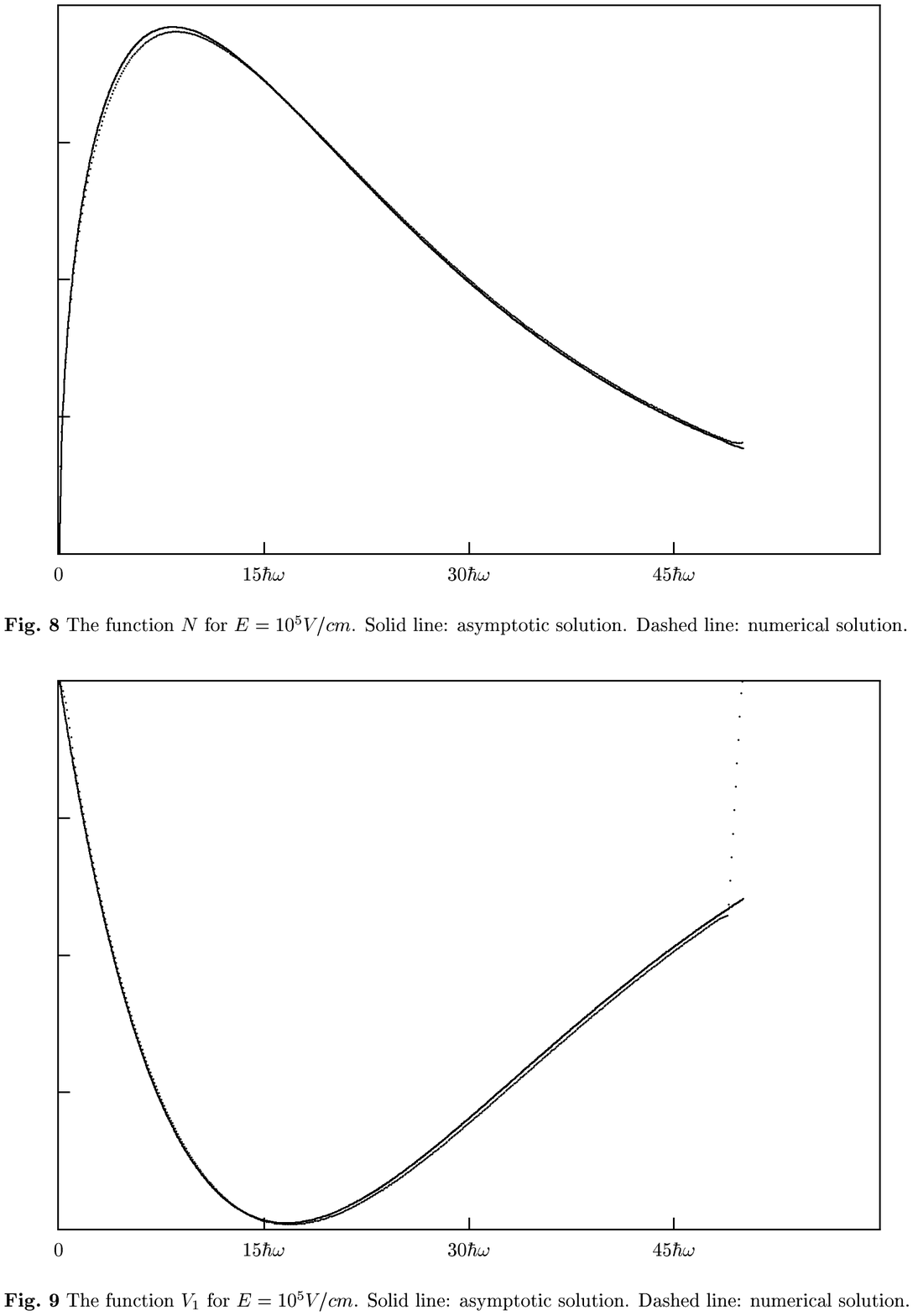}}
\end{figure}


\newpage
\small
\begin{thebibliography}{33}
%
\bibitem{Ferr} 
\auto{D.~K.}{Ferry}
\arti{Fundamental aspects of hot electron phenomena}{In: \auto{W.}{Paul} 
(ed.) Handbook on Semiconductors}{Vol.~I}{563-597 North-Holland 
Publishing Company}{1982} 
%
\bibitem{SmJe} 
\auto{H.}{Smith} and \auto{H.~H.}{Jensen}
\libr{Transport Phenomena}{New York}{Oxford Univ. Press}{1989}
%
\bibitem{VeBa} 
\auto{D.}{Ventura}, \auto{A.}{Gnudi} and \auto{G.}{Baccarani}
\arti{A Deterministic Approach to Solution of the BTE in
Semiconductors.}{Rivista del Nuovo Cimento}{18}{1-33}{1995}
%
\bibitem{Fer2} 
\auto{D.~K.}{Ferry}
\libr{Semiconductors}{New York}{Macmillan Publ. Comp.}{1991} 
%
\bibitem{VeRu} 
\auto{M.~C.}{Vecchi} and \auto{M.}{Rudan}
\arti{Modeling Electron and Hole Transport with Full-Band Structure Effects
 by means of the Spherical-Harmonics Expansion of the BTE}{IEEE Trans. 
Electron Devices}{45}{230-238}{1998}
%
\bibitem{JaLu}
\auto{C.}{Jacoboni} and \auto{P.}{Lugli} 
\libr{The Monte Carlo Method for Semiconductor Device Simulation}{New 
York}{Springer-Verlag}{1989}
%
\bibitem{Grad}
\auto{H.}{Grad}
\arti{principles of the Kinetic Theory of gases}{In: \auto{S.}{Fl\"{u}gge} 
(ed.) Handbuch der Physik, Springer-Verlag}{Vol.~12 }{205-294}{1958}
%
\bibitem{Maj3} 
\auto{A.}{Majorana} 
\arti{Spherical-Harmonics Type Expansion for the Boltzmann Equation in
Semiconductor Devices}{Le Matematiche}{LIII}{331-344}{1998}
%
\bibitem{Maj1}
\auto{A.}{Majorana} 
\arti{Conservation laws from the Boltzmann equation describing 
electron-phonon interaction in semiconductors}{Transport Theory Statist. 
Phys.}{22}{81-93}{1993}
%
\bibitem{Maj2} 
\auto{A.}{Majorana} 
\arti{Trend to Equilibrium of Electron Gas in a Semiconductor
According to the Boltzmann Equation}{Transport Theory 
Statist. Phys.}{27}{547-571}{1998}
%
\bibitem{Blak} 
\auto{J.~S.}{Blakemore} 
\libr{Semiconductor Statistics}{New York}{Dover Publ.}{1987}
%
\bibitem{RaWA} 
\auto{K.}{Rahmat}, \auto{J.}{Whithe} and \auto{D.~A.}{Antoniadis}
\arti{Simulation of Semiconductor Devices Using a
Galerkin/Spherical Harmonic Expansion Approach to Solving
the Coupled Poisson-Boltzmann System}{IEEE Trans. 
Computer-Aided Design}{15}{1181-1196}{1996}
%
\bibitem{Hans}
\auto{W.}{H\"{a}nsch} 
\libr{The Drift-Diffusion Equation and its Applications in MOSFET 
Modelling}{New York}{Springer-Verlag}{1991}
%
\bibitem{Tom} 
\auto{K.}{Tomizawa}
\libr{Numerical Simulation of Submicron Semiconductor
Devices}{Boston}{Artech House}{1993} 
%
\end{thebibliography}
\end{document}